\documentclass[10pt,a4paper]{article}

\usepackage[utf8]{inputenc}
\usepackage[T1]{fontenc}
\usepackage{amsmath,amssymb,amsfonts}
\usepackage{graphicx}
\usepackage{caption}
\usepackage{xcolor}
\usepackage{hyperref}
\usepackage{float}
\usepackage[nomarkers,nolists]{endfloat}
\usepackage{siunitx}
\usepackage{booktabs}
\usepackage{multirow}
\usepackage{makecell}
\usepackage[sort&compress,numbers]{natbib}
\usepackage{multicol}
\usepackage{authblk}

\usepackage{geometry}
\begin{document}


\title{On the importance of Ni-Au-Ga interdiffusion in the formation of a Ni-Au/p-GaN ohmic contact}

\author[1]{J. Duraz\thanks{Email: jules.duraz@cnrs.fr}}
\author[1]{H. Souissi}
\author[1]{M. Gromovyi\thanks{Present address: CRHEA, UMR7073, CNRS, Université Côte d’Azur, 06 560 Valbonne, France}}
\author[2]{D. Troadec}
\author[1]{T. Baptiste}
\author[1]{N. Findling}
\author[3]{P. Vuong}
\author[3]{R. Gujrati}
\author[3]{T.-M. Tran}
\author[3]{J.-P. Salvestrini}
\author[1]{M. Tchernycheva}
\author[3]{S. Sundaram}
\author[3,4]{A. Ougazzaden}
\author[1]{G. Patriarche}
\author[1]{S. Bouchoule\thanks{Email: sophie.bouchoule@cnrs.fr}}

\affil[1]{Centre de Nanosciences et de Nanotechnologies, C2N, UMR 9001, CNRS, Université Paris-Saclay, 91120 Palaiseau, France}
\affil[2]{Institut d’Electronique, de Microélectronique et de Nanotechnologie, IEMN, UMR 8520, CNRS, Université de Lille, 59652 Villeneuve d’Ascq, France}
\affil[3]{Georgia Tech CNRS, IRL 2958, GT-Europe, 57070 Metz, France}
\affil[4]{Georgia Institute of Technology, School of Electrical and Computer Engineering, Atlanta, GA 30332-0250, USA}

\date{} 

\maketitle

\begin{abstract}
The Ni-Au-Ga interdiffusion mechanisms taking place during rapid thermal annealing (RTA) under oxygen atmosphere of a Ni-Au/p-GaN contact are investigated by high-resolution transmission electron microscopy (HR-TEM) coupled to energy dispersive X-ray spectroscopy (EDX). It is shown that oxygen-assisted, Ni diffusion to the top surface of the metallic contact through the formation of a nickel oxide (NiO$_x$) is accompanied by Au diffusion down to the GaN surface, and by Ga out-diffusion through the GaN/metal interface. Electrical characterizations of the contact by Transmission Line Method (TLM) show that an ohmic contact is obtained as soon as a thin, Au-Ga interfacial layer is formed, even after complete diffusion of Ni or NiO$_x$ to the top surface of the contact. Our results clarify that the presence of Ni or NiO$_x$ at the interface is not the main origin of the ohmic-like behavior in such contacts. Auto-cleaning of the interface during the interdiffusion process may play a role, but TEM-EDX analysis evidences that the creation of Ga vacancies associated to the formation of a Ga-Au interfacial layer is crucial for reducing the Schottky barrier height, and maximizing the amount of current flowing through the contact.
\end{abstract}

\section{Introduction} 
GaN and related group III-nitride semiconductors are now widely used for the fabrication of opto-electronic devices such as transistors for high-power, high-temperature applications, or semiconductor lasers and light-emitting diodes (LED) in the visible-to-UV spectral range.
In such devices, a main limiting factor in the performances still is the poor quality of the ohmic contact on the p-side of the diode.
Indeed, the implementation of a low-resistive contact metallization scheme for p-GaN has represented a challenge in the past decades, due to the large activation energy of p-type dopant such as Mg ($\sim$160 – 200 meV) and hence low density of ionized acceptors typically of at most $\sim$\SI{e18}{\cm\cubed}, but also due to the lack of metals with a work-function adapted to that of p-GaN ($\sim$6.6 eV), required to minimize the barrier height of the metal/semiconductor Schottky diode\cite{Chen2015,Greco2016,Ofuonye2014,Rickert2002}.
A comprehensive analysis of the behaviour of metal/p-GaN contacts has also been complicated by variations in the structure growth conditions and devices fabrication methods, including the control of the GaN/metal interface, leading among others to important variations in the Fermi level pinning effect, impacting the electrical performance of the contact\cite{Rickert2002,Wahid2020}.
Nevertheless, high-work-function ($\sim$5 eV) metals such as Ni, Au, Pt, Pd have been mostly studied in order to minimize the Schottky barrier height, and to achieve an ohmic behaviour\cite{Greco2016,Cho2005,Lee1999,Jang2000,Zhou2000,Huh2001}.
Thermal annealing of Ni-Au/p-GaN contact was first investigated to reduce the specific contact resistance\cite{Sheu1998}, and a beneficial effect of annealing under oxygen-containing atmosphere was discovered\cite{Ho1999,Mistele2001}, attributed to the formation of a NiO/p-GaN interface\cite{Ho1999,Chen1999}.
However, other works concluded that NiO was not the reason for the Ni-Au metallization to yield ohmic behaviour after thermal annealing in an oxidizing atmosphere\cite{Maeda1999}.
Others effects were thus emphasized to explain the ohmic-like behaviour after annealing, namely the removal of surface contamination or native oxide during Ni/Au layers reversal occurring during annealing\cite{Qiao2000,Koide1999,Lee1999}, or the creation of Ga vacancies in p-GaN close to the metal interface\cite{Jang2003,Kim2000,Sarkar2018}.
Some authors pointed out the formation of an (Ni-)Au-Ga alloy in contact with p-GaN\cite{Qiao2000,Jang2003,Kim2000,Sarkar2018}, possibly involved in the reduction of the Schottky barrier height.
However, the exact role, and relative importance of the different mechanisms mentioned above has not been clarified yet.
In a recent experimental study using TEM coupled to EDX analysis, it was again supposed that the creation of Ga vacancies may play a role, but the presence of a NiO/p-GaN interface was inferred to have a major importance in the formation of an ohmic contact\cite{Mauduit2023}.

We have conducted a detailed TEM-EDX study attempting to elucidate the exact role of Ni/NiO$_x$, Au, or (Ni-)Au-Ga alloy in contact with p-GaN in the formation of an ohmic-like p-GaN contact, and to clarify the importance of the different diffusion processes assisted by thermal annealing under an oxygen atmosphere.
Using the same epitaxial structure, we have compared the electrical performance of the p-GaN/Ni-Au contact for different annealing conditions, and different metal evaporation conditions, and we have correlated the electrical results with high-resolution TEM coupled to EDX analysis.
From our results, we conclude that the creation of Ga vacancies during Ni-Au/Ga interdiffusion processes plays the major role in the reduction of the Schottky barrier height.

\section{Experimental conditions and methods}

\begin{figure}
    \centering
    \includegraphics[width=10cm]{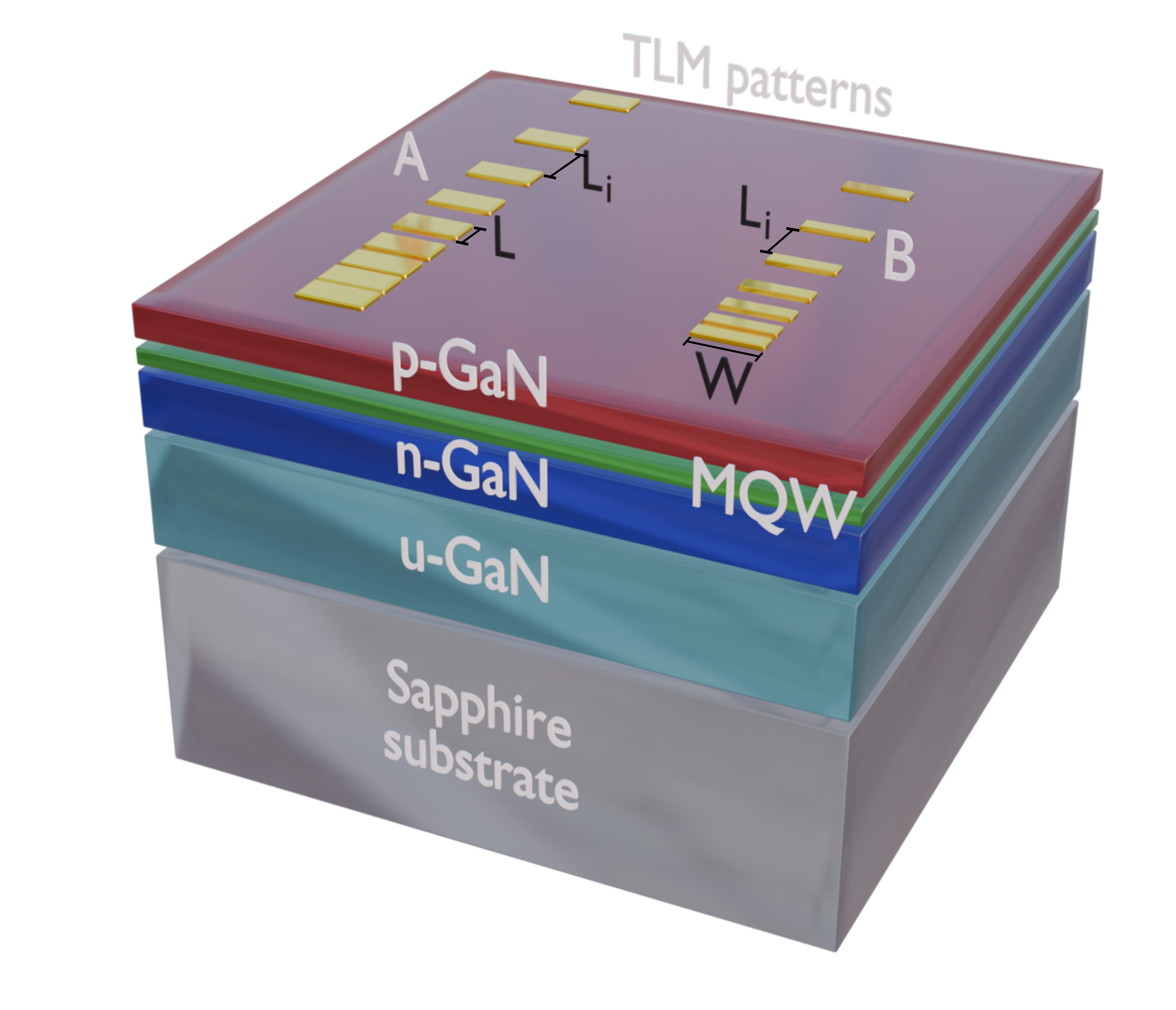}
    \caption{Schematics of the epitaxial structure used for the study, and geometry of the TLM pads used for the electrical characterization of the contact. $W$ = 200 µm is the pad width. $L$ = 100 µm (A) or 50 µm (B) is the TLM pad length, with (A) $L_i$ = 5/ 10/ 20/ 40/ 70/ 100/ 150/ 200 µm and (B) $L_i$ = 20/ 40/ 60/ 100/ 150/ 200 µm, the distance between two successive pads.}
    \label{fig:sample}
\end{figure}

The epitaxial structure used for the study is a standard InGaN/GaN-based blue LED MQW heterostructure emitting at 460 nm grown by MOCVD on a 2-inch c-plane (0001) sapphire substrate, schematically depicted in Figure~\ref{fig:sample}.
It consists of a 1-µm thick n-doped GaN layer grown on a 2-µm thick unintentionally doped GaN buffer layer, an active region comprising five InGaN quantum wells with a thickness of 1.8 nm ± 0.2 nm separated by 12-nm thick GaN barriers, and a Mg-doped, 270-nm thick p-GaN layer terminated by a 12-nm thick p$^+$-GaN surface layer.
The Mg concentration is typically of \SI{1e19}{\per\cm\cubed}, and the average p-dopant concentration is estimated to be of \SI{1e17}{\per\cm\cubed} in the p-GaN layers after dopant activation and may reach up to \SI{3e17}{\per\cm\cubed} in the p$^+$-GaN surface layer.

The wafer with its surface covered by a protective resist layer has been diced into samples with a surface of 10 mm × 10 mm.
After dicing, the samples are first cleaned in acetone and isopropanol (IPA), then dipped for 10 min in a piranha-like solution (H$_2$SO$_4$ – 20 vol. / H$_2$O$_2$ – 6 vol. / H$_2$O - 25 vol.) and rinsed for 3 min in deionised water (DIW) in order to remove any trace of organic residues.
The samples are finally dipped for 3 min into a HCl-based solution (2 vol. of DIW / 1 vol of 37\% HCl) and rinsed into DIW prior to a UV negative photolithography step (AZ5214 photoresist in reversal mode, and TMAH-based developer AZ 826 MIF), to define the TLM patterns described in Figure~\ref{fig:sample} used for the electrical characterization of the p-GaN/metal contacts.
An ex-situ deoxidation sequence is applied prior to metal evaporation, consisting of the same HCl-based solution as the one used before the lithography step, completed by a dip for 1 min in 10\% HF-based solution followed by a short rinse into DIW, and N$_2$ drying.
It was observed separately from ex-situ XPS analysis of the GaN surface, that this chemical deoxidation sequence minimized the amount of native gallium oxide (GaO$_x$) at the GaN surface, which residual thickness was estimated to be lower than $\sim$0.3 nm within the 10 minutes following the deoxidation.
The sample is then immediately installed into the loadlock of the electron-beam vacuum evaporator, which is pumped down to a pressure of \num{5e-7} mbar before the transfer of the sample into the evaporation chamber, maintained at a base pressure in the range of 1-\num{2E-8} mbar by cryo-pumping.
A 20-nm thick Ni layer, and a 200-nm thick Au top layer, are deposited at a constant evaporation rate of 0.5 nm/s, regulated by a quartz oscillator.
The typical mass of solid Ni, and Au metals installed in the 15 cm$^3$ Cu crucibles of the evaporator lie in the range of 100 g to 150 g typically, and the crucible-to-sample distance is of 65 cm leading to an e-beam current of 220 mA, and 420 mA, during Ni, and Au evaporation at 0.5 nm/s, respectively.
The pressure is measured to be of \num{2E-8} mBar / \num{1E-7} mBar during Ni evaporation and may increase up to \num{7E-8} mBar / \num{3E-7} mbar during evaporation of the thicker Au layer, at the crucible position / at the sample position, respectively.
The TLM patterns are finally defined after lift-off in appropriate solvent bath (acetone, and NMP-containing stripper Remover PG), followed by an IPA rinse and N$_2$ drying.

Except for the non-annealed samples, rapid thermal annealing (RTA) of the contacts is performed at atmospheric pressure under pure oxygen ambient.
The samples are placed into the oven (JIPELEC JetFirst) onto a SiC-coated, graphite plate covered by a SiC-coated graphite cover.
The oven is pumped down to few mBars then purged with oxygen twice before starting the RTA sequence with a constant oxygen flow fixed to 500 standard cubic centimetre per minute (sccm).
Heating is ensured by halogen lamps placed at a distance of $\sim$45 mm above the cover.
A thermocouple fixed to the backside of the graphite plate is used to regulate the annealing temperature, which is generally fixed to \SI{450}{\celsius} for annealing times varied from 2 min to 10 min in the experiments, except for some tests performed at \SI{350}{\celsius}, and up to \SI{650}{\celsius}.
At the end of annealing time and after halogen lamps switch-off, the oxygen flow is stopped and the oven is pumped to vacuum for typically 30 s until the temperature has decreased by at least \SI{100}{\celsius}.
Nitrogen flow is then switched on to accelerate the cooling down to a temperature of \SI{100}{\celsius} before removing the sample from the oven.

The TLM approach is used to characterize the electrical contacts\cite{Reeves1982}.
As depicted in Figure~\ref{fig:sample}, the distance $L_i$ between TLM pads can be varied from 5 µm to 200 µm.
Two-probes measurements are performed to retrieve the I-V characteristic between two successive TLM pads in the [-5 V; +5 V] voltage range.
The overall resistance $R$ is determined from a linear fit of the characteristic in the [4 V – 5 V] voltage interval.
The TLM pad contact resistance, $R_c$, is deduced from the linear fit of the curve $R$($L_i$) with $L_i$ varying in the range from 5 µm to 70 µm.
The transfer length $L_T$, the contact specific resistance $r_c$, and the semiconductor sheet resistance $R_{sh}$ are deduced following the TLM method\cite{Reeves1982}.
An estimate of the Schottky barrier height, $\Phi_B$, is obtained by fitting the I-V characteristic in the [-1 V ; 0 V] voltage range for which one SC/metal diode is reversed-biased, following the approach of Mori et al. or Wahid et al.\cite{Mori1996,Wahid2020}.
Considering the targeted application of a p-contact in LEDs or laser diodes, the contacts are qualified by a complementary figure of merit, corresponding to the current $I_{\text{max}}$ flowing between two TLM pads separated by a distance $L_i$ = 20 µm, at a voltage difference set to 5 V.
Indeed, the maximum of current driven by the contact may be practically considered as the most relevant figure of merit\cite{Tung2014}.
The values of $r_c$ (\unit{\ohm\cm\squared}), $I_{\text{max}}$ (mA), and $\Phi_B$ (eV) are listed in Table~\ref{tab:sample-results} for each sample of this study.
The typical value of the sheet resistance of the p-GaN layer, $R_{sh}$, does not vary significantly from sample to sample and is found to be in the range of (3 ± 0.5)$\times$ \num{e4} $\Omega/\Box$.
The $L_T$ value is found to be always smaller than 15 µm, leading to similar I-V characteristics and contact resistance results, for the TLM pad length L = 50 µm, or L = 100 µm.
For comparison purposes, the TLM I-V characteristics of the different samples are exemplified for a fixed distance $L_i$ = 20 µm in the following.

After electrical characterization, part of each sample is coated by a $\sim$70-nm thick a-C layer, and a TEM slice with a typical width of 15 µm and height of 3 µm is taken from a TLM pad in the a-C coated area by Ga-Focused-Ion-Beam (FIB) etching, for further physico-chemical and structural analysis of the GaN – metal stack and GaN/metal interfacial layer.
The a-C layer is completed by a local deposition of Pt in the FIB chamber.
Several 2-µm wide windows are finely FIB-etched in the slice down to a thickness typically varying from 200 nm to slightly less than 100 nm.
A specific fine etching step terminates FIB etching to minimize Ga redeposition at the facets.
Thin windows are preferentially used for HR-TEM observations while thicker ones are preferentially used for TEM-EDX analysis.
EDX coupled to TEM allows to estimate the atomic composition of the analysed area in different ways.
Firstly, 2D cartographies of the relative distribution of the different atoms present in the analysed area can be reconstructed, referred to as EDX map in the following.
Secondly a quantitative estimation of the local composition in atomic percentage can be estimated from the EDX spectra: the composition in atomic percentage can be calculated at several points along a linescan defined perpendicular to the layer interfaces, leading to a compositional profile.
However, in the linescan option, the quantitative atomic composition is determined from the EDX spectrum at each point, considering a TEM slice thickness, $d$, a material density $\rho$, and hence the $\rho \times d$ product, as constant along the linescan.
This may not be the case, since for example the theoretical density of pure gold is $\sim$19 g/cm$^3$, while that of Ni is of $\sim$9 g/cm$^3$, and that of GaN is rather 6 g/cm$^3$, close to that of NiO.
The slide thickness may also slightly locally vary, especially at the interfaces between two different layers.
It was checked during the quantitative analysis of a linescan, that for a slice thickness $d$ in the range between 100 nm to 200 nm, setting the material density from 15 g/cm$^3$ down to 5 g/cm$^3$, leads to a difference in the final calculated atomic percentage lower than 3\%.
This will be considered as the precision limit of the atomic percentage values in the following.
As an alternative to the linescan, several rectangular zones can be manually selected in homogeneous regions of the analysed area, and the quantitative, average composition in atomic percentage can be calculated in each zone independently.
In this way, the material density can be fixed to a value consistent with the composition observed from the average EDX spectrum in each zone.
Finally, for a correct consideration of the results presented in the next sections, it should be noted that two physical effects may limit the precision of the calculated atomic percentages from the EDX spectra.
Firstly, the correct determination of the atomic percentage of light atoms such as oxygen is difficult in regions or points where heavy atoms such as Au are predominant, due to the strong, and broadband Bremsstrahlung effect.
It may be interpreted in the spectrum automatic fitting as a signal for the light atom, leading to an overestimation of the latter.
The atomic percentage of oxygen will therefore be overestimated in gold-rich zones.
Secondly, atoms may be present only at the surface of the TEM slice.
Despite the FIB final fine etch, Ga atoms may be present, leading to an overestimation of the Ga percentage.
The possible Ga pollution can be estimated by analysing a zone in the a-C coating for example.
It is found that this pollution never contributes to more than 2-3\% of Ga in the calculated percentage for a slice thickness in the range 100 nm – 200 nm.
The slice facets are also expected to be oxidized over a depth of typically 1 nm.
The resulting signal in the EDX spectrum, originating from the surface, is interpreted in the spectrum fitting as a signal from the volume of the slice.
This leads to an overestimation of light atoms, and hence an overestimation of oxygen.
Thicker windows in the TEM slice are therefore preferred for TEM-EDX to minimize the oxygen signal originating from the surface, compared to the oxygen signal from the volume.

\section{Experimental results and analysis}

\begin{figure}
    \centering{
    \includegraphics[width=10cm]{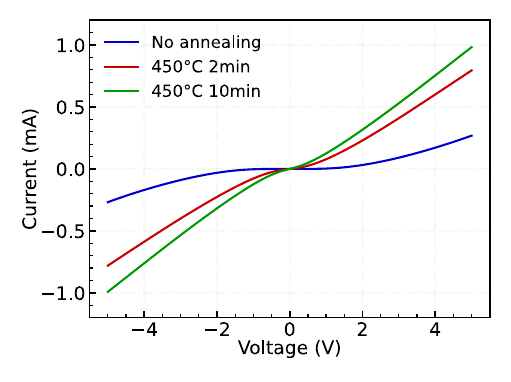}} 
    \caption{TLM I-V characteristics measured for a distance between pads Li = 20 µm without annealing (blue curve), after an-nealing for 2min at 450°C under O2 atmosphere (red curve), and after annealing for 10min at 450°C under O2 atmos-phere (green curve). The initial thicknesses of the Ni, and Au layers evaporated at a rate of 0.5 nm/s are of 20nm, and 200nm, respectively.}
    \label{fig:TLM}
\end{figure}

Figure~\ref{fig:TLM} shows the TLM I-V curve of the non-annealed sample, for a distance between TLM pads $L_i$ = 20 µm.
A Schottky barrier is clearly apparent, estimated to be of $\sim$0.70 eV in height as reported in Table~\ref{tab:sample-results}, and a rather low $I_{\text{max}}$ value of 0.30 mA is reached at 5 V.
The specific contact resistance value is found to be $r_c$ $\sim$ \SI{1.5e-1}{\ohm\cm\squared}.
The I-V curve presents a more ohmic-like profile after RTA for 2 min at \SI{450}{\celsius} under O$_2$ atmosphere.
The $I_{\text{max}}$ value is notably increased (to $\sim$0.80 mA), associated to the reduction of Schottky barrier height, and the specific contact resistance is decreased down to \SI{7.5e-3}{\ohm\cm\squared}.
A longer annealing time of 10 min slightly improves the results, however $I_{\text{max}}$ tends to saturate to $\sim$1 mA, and $r_c$ is only moderately reduced (down to \SI{3.7e-3}{\ohm\cm\squared}).

\begin{table*}
\centering
\caption{Summary of results for samples with different Ni/Au thickness and annealing conditions.}
\setlength{\tabcolsep}{4pt}
\begin{tabular}{cccccc}
\toprule
Sample & Ni/Au thickness$^a$ & Annealing conditions & $I_\text{max}$ (mA) & $\Phi_B$ (eV) & $r_c$ ($\Omega\cdot$cm$^2$) \\
\midrule
A & 20/200 & No annealing & 0.30 & 0.68 & $1.5 \times 10^{-1}$ \\
B & 20/200 & \SI{450}{\celsius}, O$_2$, 2 min & 0.78 & 0.52 & $7.5 \times 10^{-3}$ \\
C & 20/200 & \SI{450}{\celsius}, O$_2$, 10 min & 0.98 & 0.49 & $3.7 \times 10^{-3}$ \\
D & 3/3     & No annealing & 0.60 & 0.55 & $1.3 \times 10^{-2}$ \\
E & 3/3     & \SI{450}{\celsius}, O$_2$, 2 min & 0.99 & 0.46 & $1.6 \times 10^{-2}$ \\
F & 3/3     & \SI{350}{\celsius}, O$_2$, 2 min & 0.77 & 0.50 & $2.8 \times 10^{-2}$ \\
G & 3/3     & \SI{350}{\celsius}, O$_2$, 10 min & 0.88 & 0.47 & $3.1 \times 10^{-2}$ \\
H & 3/3 + 20/50$^b$ & \SI{650}{\celsius}, O$_2$, 2 min & 0.85 & 0.53 & $2.0 \times 10^{-2}$ \\
\bottomrule
\end{tabular}
\par\vspace{0.5em} 
$^a$The evaporation rate of the thick metal layers (Ni-20nm / Au-200nm) is of 0.5 nm/s, and of 0.1 nm/s for the thin metal layers (Ni-3nm / Au-3nm). $^b$A second Ni-Au evaporation on the TLM pads subsequent to thermal annealing has been carried out on sample H to allow for the electrical measurements.
\label{tab:sample-results}
\end{table*}

\begin{figure}
\centering{\includegraphics[width=\linewidth]{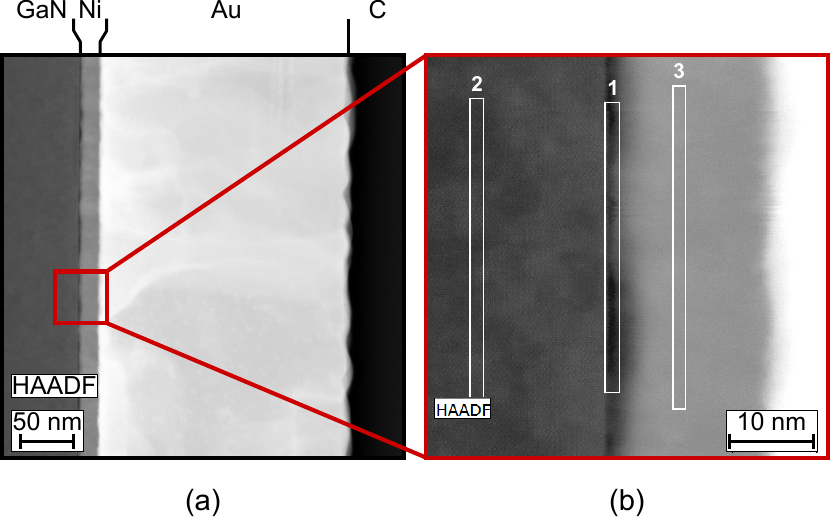}} 
  \caption{
  (a) - HAADF cross-sectional STEM image of the GaN-Ni-Au stack for the non-annealed sample. The metal layers evapo-rated at a rate of 0.5 nm/s have a nominal and measured thickness of 20 nm, and 200 nm, for Ni, and Au, respectively. (b) - HAADF STEM image at higher magnification of the region marked with a red square in (a). The three white, 35-nm long, 1.5-nm wide rectangles labelled as 1 (correspond-ing to the interfacial layer between GaN and Ni), 2 (located in GaN, 15 nm below the GaN/Ni interface), and 3 (located in Ni, 8 nm above the GaN/Ni interface) correspond to the three zones for the determination of the quantitative average composition in atomic percentage listed in Table~\ref{tab:NR}.}
  \label{fig:NoAnnealing}
\end{figure}

TEM-EDX analyses have been carried out for the non-annealed sample, as well as after annealing for 2 min, and for 10 min.
Figure~\ref{fig:NoAnnealing}-(a) shows a High-Angle Annular Dark-Field (HAADF) cross-sectional HR-STEM image of the GaN- Ni-Au stack for the non-annealed sample.
It is observed that the starting of Ni deposition on the GaN surface is not completely uniform, with nano-grains formed in the Ni layer, leading to a rms surface roughness of $\sim$1.4 nm estimated from the analysis of the HAADF image Z-contrast, assuming a gaussian distribution.

The 200-nm thick gold layer also shows some larger nano-grains, leading to an estimated top surface rms roughness of $\sim$2 nm in the image.
The GaN/Ni interface shows a lower intensity on the HAADF STEM image, which may be interpreted either as nano-voids embedded in the TEM slice thickness, or as the presence of a lighter material, presumably an oxide.
The average thickness of this interfacial layer is estimated to be 2.9 nm, from the analysis of the average Z-contrast in the HAADF-STEM image of Fig.~\ref{fig:NoAnnealing}-(a).
A quantitative analysis of the average atomic composition of the interfacial layer is performed in the selected zone labelled as \#1 in Figure~\ref{fig:NoAnnealing}-(b).
Ga, N, Ni, and O can be detected in the average EDX spectrum, while Au is not detected.
The atomic composition of the interface is compared to that of zones labelled as \#2, and \#3, in the GaN, and Ni layers, respectively.
The results are listed in Table~\ref{tab:NR}.
A 50/50 GaN stoichiometry is reasonably retrieved in zone \#2, with the presence of a low amount of oxygen, close to the noise level ($\sim$1-2\%); as discussed previously, this signal probably originates from the native gallium oxide at the TEM slice facets.
Ni oxidation is much less pronounced in zone \#3, with oxygen at the noise level in the corresponding average EDX spectrum.
Almost no trace of Au, nor of Ga, can be found in the 20-nm thick Ni layer, which acts as a diffusion barrier.
The Ga atomic percentage of 1.9\% is very close to the $\sim$2\% value generally retrieved and attributed to Ga FIB pollution at the slice facets.

In contrast with zone \#2 and \#3, a more important oxidation can be measured in the interfacial layer (zone \#1), with an oxygen percentage of 8.5\% ± 1\% (depending on the $\rho \times d$ product value).
The interfacial layer therefore corresponds to a presumably discontinuous, GaNiO$_x$ layer.
It may explain the presence of the Schottky barrier observed in Figure~\ref{fig:TLM} and Table~\ref{tab:sample-results}, but may also induce some Fermi level pinning effect impacting the effective Schottky barrier height\cite{Tung2014,Wahid2020}.

\begin{table}
\centering
\caption{Average atomic percentage in each rectangular zone.}
\label{tab:NR}
\setlength{\tabcolsep}{4pt}
\begin{tabular}{lccc}
\toprule
\multirow{2}{*}{Elements} & \multicolumn{3}{c}{\makecell{Average atomic percentage in each\\rectangular zone}} \\
\cmidrule{2-4}
 & \#1 (interfacial)$^a$ & \#2 (GaN)$^b$ & \#3 (Ni)$^c$ \\
\midrule
Ga & 47.4 & 49.3 & 1.7 \\
N  & 3.9  & 47.4 & $\varnothing$ \\
Ni & 40.2 & $\varnothing$    & 98.3 \\
O  & 8.5  & 3.3  & $\varnothing$ \\
\bottomrule
\end{tabular}
\par\vspace{0.5em}
The TEM slice thickness is fixed to 200 nm and the average material density to $^a$6 g/cm$^3$, $^b$5 gm/cm$^3$ and $^c$8 g/cm$^3$. The $\varnothing$ symbol represent an absence of detection of the element.
\end{table}

\begin{figure}
\centering{\includegraphics[width=12cm]{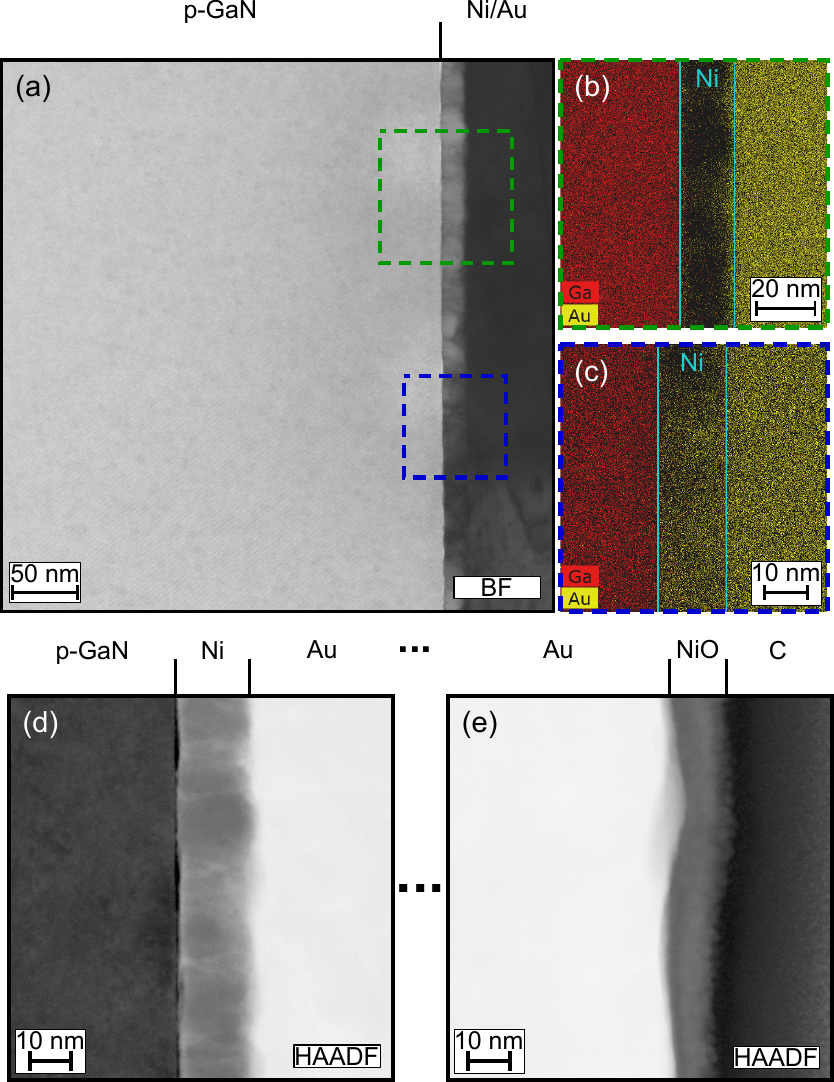}} 
  \caption{
  Cross-sectional HR-STEM images and EDX mapping showing the relative distribution of Au and Ga atoms for the sample after RTA for 2min at 450°C under O2 atmosphere. (a): Large view, Bright-Field (BF) STEM image of the p-GaN, Ni, Au stack. (b): EDX map in a region where Au interdiffusion is moderate; (c): EDX map in a region where Au interdiffusion is strong; (d): HAADF HR-STEM image at high magnification of the GaN/Ni/Au layers; (e): HAADF HR-STEM image at high magnification of the Au/NiOx top surface.}
  \label{fig:2min}
\end{figure}

Figure~\ref{fig:2min}-(a) shows a large-view, cross-sectional Bright-Field (BF) HR-STEM image of the GaN/Ni/Au layers after RTA for 2 min at \SI{450}{\celsius} in O$_2$ ambient.
The structure of the initial Ni layer is strongly modified by the annealing, with the apparition of polycrystalline nano-domains of high, and low intensity, to be interpreted as low-density, and high-density materials, respectively.
Figure~\ref{fig:2min}-(b) corresponds to the EDX map of a high-intensity area marked with a green dashed square in Fig.~\ref{fig:2min}-(a).
Ni is still present, however Au interdiffusion in Ni is taking place through channels that may be related to the nano-grain boundaries in the initial Ni layer.
Moreover, some Au accumulation is apparent within $\sim$2–3 nm right above the GaN surface.
Traces of Ga (signal above the Ga K-line noise level in the EDX spectrum) are also detectable in the initial Ni layer, as well as in the initial Au layer, on the EDX maps; however, quantitative analysis of the atomic composition is required to evaluate more precisely the importance of Ga diffusion.
Fig.~\ref{fig:2min}-(c) corresponds to the EDX map of a low-intensity area marked with a blue dashed square in Fig.~\ref{fig:2min}-(a).
Ni is still present, however Au interdiffusion is much more pronounced in this area, which is consistent with a higher-density metallic alloy.
A dark line within 1–2 nm just above the GaN surface can be noticed in the two EDX maps, which indicates that part of the GaNiO$_x$ layer identified before annealing is still present, with some Au incorporation added.
Indeed, weak traces of O are detectable close to the GaN surface.
Quantitative analysis of the atomic composition is required to evaluate more precisely the amount of oxygen in the metallic alloy.
As a first conclusion, the reduction of the Schottky barrier height observed in Fig.~\ref{fig:TLM} after annealing for 2 min may be associated to the diffusion processes taking place in Fig.~\ref{fig:2min}.
As previously introduced, three physical effects may be involved : firstly a reconstruction or “auto-cleaning” of the GaN/metal interface occurring during the interdiffusion\cite{Lee1999,Qiao2000,Koide1999} removing impurities and reducing the thickness of the native GaO$_x$ oxide ; secondly the presence of Au close to the GaN surface, and presumably of a (Ni-)Ga-Au alloy formed with the observed Ga diffusion in the metallic layer\cite{Qiao2000,Kim2000,Sarkar2018}, leaving Ga vacancies at the GaN surface\cite{Jang2003,Kim2000,Sarkar2018} ; thirdly, the presence of a NiO$_x$ layer in contact with the GaN surface\cite{Ho1999,Chen1999,Mauduit2023}. 

Figure~\ref{fig:2min}-(d) shows a high-magnification, cross-sectional HAADF HR-STEM image of the GaN/Ni/Au stack.
The Z-contrast evidences that the low-density interfacial layer detected before annealing is still present, but restricted to some discontinuous areas (dark areas) at the nano-scale.
The channels of Au diffusion into the initial Ni layer can also be identified in the Z-contrast HAADF image, as brighter zones in the layer Fig.~\ref{fig:2min}-(e) shows a cross-sectional HAADF HR-STEM image of the Au top surface at a high magnification.
A new layer has been formed on top of the 200-nm thick layer, consisting of NiO as deduced from EDX map and analysis (see Table~\ref{tab:2min}); it can be observed almost everywhere on top of gold along the TEM slice.
The average quantitative composition in atomic percentage can be calculated with the same approach as that used in Fig.~\ref{fig:TLM}-(b).
Table~\ref{tab:2min} summarizes the results obtained for several rectangular zones selected in the EDX maps associated to Fig.~\ref{fig:2min}-(b), Fig.~\ref{fig:2min}-(c), and Fig.~\ref{fig:2min}-(e), corresponding to the GaN/Ni/Au stack with moderate Au interdiffusion, the GaN/Ni/Au stack with strong Au interdiffusion, and the top Au/NiO$_x$ layer, respectively.
The quantitative analysis in zones of the EDX map corresponding to Fig.~\ref{fig:2min}-(e) shows that the top amorphous NiO$_x$ layer is oxygen-rich with a retrieved Ni/O ratio close to 40/60.
The top part of the initial Au layer remains mainly composed of gold ($\sim$95\% in atomic percentage) and Ni is almost not present, consistently with a fast Ni out-diffusion and accumulation at the surface after oxidation.
The Ga traces in these zones can be attributed to Ga FIB pollution at the slice facets.
The quantitative analysis in zones \#1 in the EDX maps corresponding to Fig.~\ref{fig:2min}-(b), and Fig.~\ref{fig:2min}-(c), shows that the bottom part of the Au layer close to the initial Ni/Au interface, also contains a moderate amount of Ni ($\sim$6\%), while traces of Ga may be attributed to FIB pollution.

The main effect of the thermal annealing can be observed in the initial Ni layer; its composition is non-uniform, as already evidenced by the Z-contrast HAADF-STEM images and EDX maps in Fig.~\ref{fig:2min} (a)-(c).
In regions such as that of Fig.~\ref{fig:2min}-(b), where Au diffusion into the Ni initial layer is moderate, an Au atomic percentage of $\sim$5\% is calculated in average (zone \#2 of the EDX map located $\sim$10 nm above the GaN/Ni interface – Fig.~\ref{fig:2min}-(b)), and up to $\sim$9\% locally (zone \#3 in the EDX map – Fig.~\ref{fig:2min}(b)).
The situation is different closer to the GaN/Ni interface, where Au incorporation reaches $\sim$14\% in average (zone \#4 in the EDX map, 0.7-nm wide, located $\sim$1.4 nm above the GaN surface – Fig.~\ref{fig:2min}-(b)), confirming quantitatively the Au interdiffusion and accumulation at the GaN interface, qualitatively deduced from the relative distribution of Au in the EDX map of Fig.~\ref{fig:2min}-(b).
The presence of Ga at an atomic percentage of $\sim$8\%, much larger than the value attributed to Ga FIB pollution at the facets, is the most striking result of the quantitative analysis in zone \#4 where Au accumulation is observed.
This indicates that Ga out-diffusion from GaN has started, thereby creating Ga vacancies at the top GaN surface.

This effect is confirmed by the quantitative analysis in zones \#2 to \#4 selected in the Ni initial layer in the EDX map corresponding to Fig.~\ref{fig:2min}-(c).
Au diffusion into the Ni initial layer is strong in this region, with Au incorporation up to 40\% in zone \#2 located $\sim$12 nm above the GaN/Ni.
At this distance, the Ga (3.5\%) incorporation is slightly larger than that expected from Ga FIB pollution at the slice facets, but still at a low level.
In contrast, Ga incorporation is larger than 6\% and up to 14\% in zones \#3 and 4 closer to the GaN/metal interface (at $\sim$3 nm from the interface).
This level of Ga incorporation implies a Ga out-diffusion from the GaN starting during annealing, especially in regions where Au accumulation is strong.

It can be concluded that annealing under O$_2$ ambient activates Ni diffusion up to the surface, where it is instantaneously oxidized.
The Ni diffusion process takes place through the 200-nm thick Au layer for annealing times as short as 2 min.
Ni constant oxidation at the surface seems to be crucial in promoting the diffusion process; indeed, the same thermal annealing procedure at \SI{450}{\celsius} repeated under N$_2$ ambient for annealing times from 2 min to 10 min, does not improve the I-V characteristics of the contact, and Ni diffusion has not been detected on EDX maps and cross-sectional HR-STEM images (not shown).
Ni diffusion is accompanied by Au diffusion into the initial Ni layer down to the GaN surface, thereby possibly reconstructing or “auto-cleaning” the GaN/metal interface.
More importantly, Ga out-diffusion from GaN is clearly ignited after 2 min annealing and seems to be more pronounced in regions where Au accumulation is strong, creating Ga vacancies in the p-GaN.

\begin{table}
\centering
\caption{Average composition calculated in the zones selected in Figure 4-(a), (b), and (e) for the sample annealed for 2 min at \SI{450}{\celsius} in O$_2$ ambient.}
\renewcommand{\arraystretch}{1.8}
\setlength{\tabcolsep}{4pt}
\begin{tabular}{ccccccc}
\hline
\makecell{Corresponding\\Figure} & \makecell{Rectangular boxes\\positions located in\\ the EDX map} & \makecell{Element} & \multicolumn{4}{c}{\makecell{Average atomic\\ percentage (\%) in each\\rectangular box \#}} \\
\hline
\multirow{5}{*}{Figure~\ref{fig:2min}-(b)} & \multirow{5}{*}{\raisebox{-0.5\height}{\includegraphics[height=3.5cm]{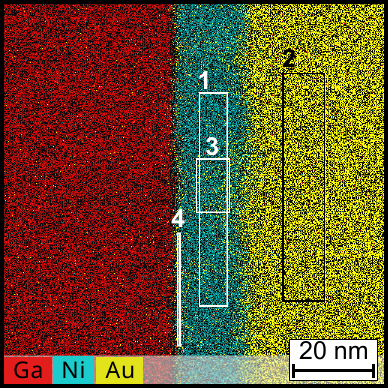}}} & & \#1 & \#2 & \#3 & \#4 \\
\cline{4-7}
& & \multirow{2}{*}{Ni} & \multirow{2}{*}{6.3} & \multirow{2}{*}{93.5} & \multirow{2}{*}{89.5} & \multirow{2}{*}{77.3} \\ 
& & \multirow{2}{*}{Au} & \multirow{2}{*}{92.2} & \multirow{2}{*}{5.0} & \multirow{2}{*}{9.0} & \multirow{2}{*}{14.0} \\ 
& & \multirow{2}{*}{Ga} & \multirow{2}{*}{1.5} & \multirow{2}{*}{1.5} & \multirow{2}{*}{1.5} & \multirow{2}{*}{8.1} \\ 
& & & & & & \\
\hline
\multirow{5}{*}{Figure~\ref{fig:2min}-(c)} & \multirow{5}{*}{\raisebox{-0.5\height}{\includegraphics[height=3.5cm]{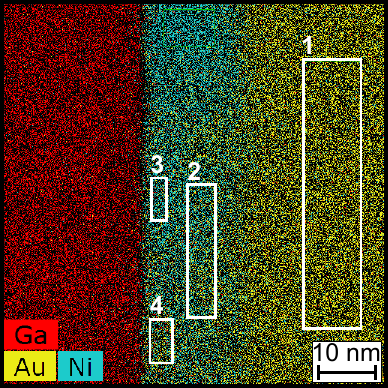}}} & & \#1 & \#2 & \#3 & \#4 \\
\cline{4-7}
& & \multirow{2}{*}{Ni} & \multirow{2}{*}{6.8} & \multirow{2}{*}{54.5} & \multirow{2}{*}{72.5} & \multirow{2}{*}{41.3} \\ 
& & \multirow{2}{*}{Au} & \multirow{2}{*}{91.3} & \multirow{2}{*}{42.0} & \multirow{2}{*}{20.8} & \multirow{2}{*}{45.2} \\ 
& & \multirow{2}{*}{Ga} & \multirow{2}{*}{1.9} & \multirow{2}{*}{3.5} & \multirow{2}{*}{6.7} & \multirow{2}{*}{13.5} \\ 
& & & & & & \\
\hline
\multirow{5}{*}{Figure~\ref{fig:2min}-(e)} & \multirow{5}{*}{\raisebox{-0.5\height}{\includegraphics[height=3.5cm]{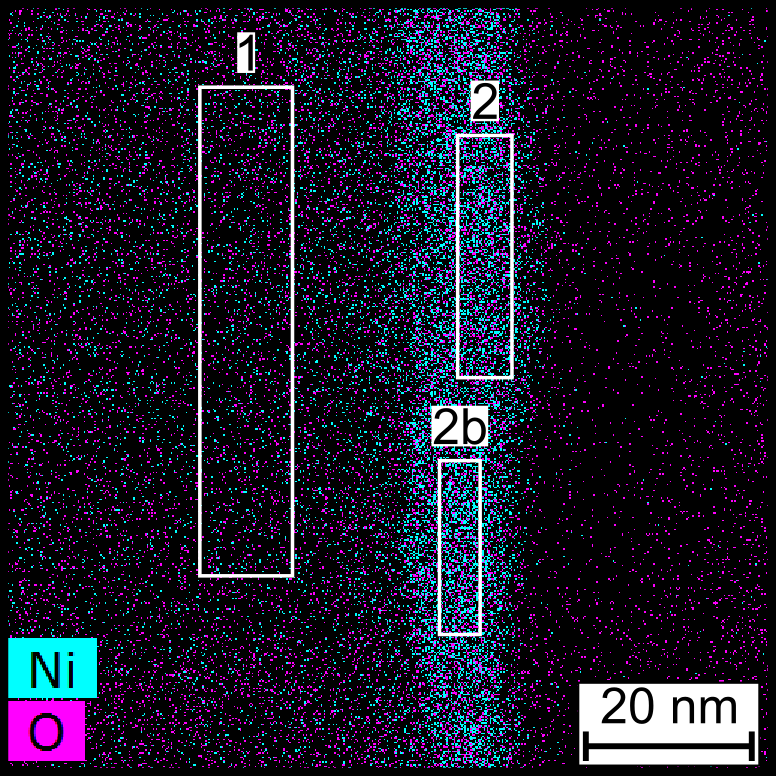}}} & & \#1 & \#2+2b & & \\
\cline{4-5}
& & Ni & 3.9 & 40.8 & & \\ 
& & Au & 94.9 & $\varnothing$ & \\ 
& & Ga & 1.2 & 1.4 & & \\ 
& & O & $\varnothing$ & 57.8 & & \\ 
\hline
\end{tabular}
\par\vspace{0.5em}
The TEM slice thickness is fixed to 100 nm, and the value of material density fixed in each zone is of 19 g/cm$^3$ for Au-rich zone, 9 g/cm$^3$ in Ni-rich zones, and 6 g/cm$^3$ in the NiO$_x$ top layer.
\label{tab:2min}
\end{table}

\begin{figure}
\centering{\includegraphics[width=\linewidth]{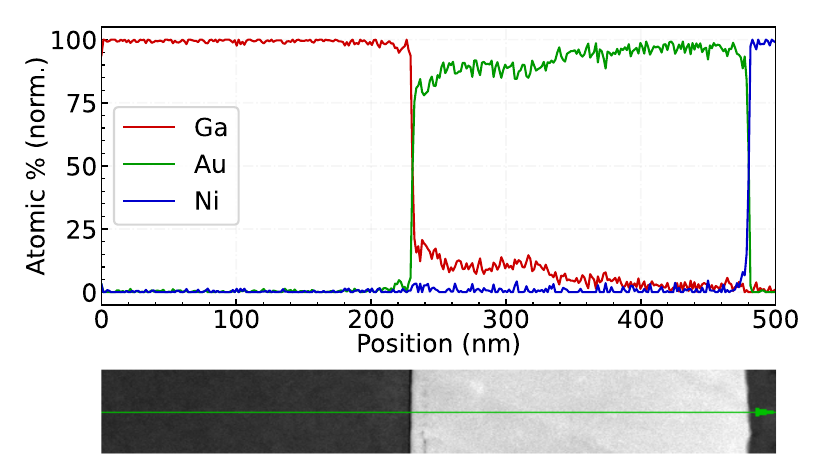}} 
  \caption{
  Relative compositional profile in atomic percentage for Ga (red curve), Au (green curve), and Ni (blue curve) retrieved from EDX spectra fitting along the linescan highlighted as a green line in the bottom, HAADF-STEM image.}
  \label{fig:Linescan}
\end{figure}

Figure~\ref{fig:Linescan} shows an EDX linescan performed in the direction perpendicular to the layer interfaces on a HAADF-STEM image, after RTA for 10 min at \SI{450}{\celsius} under O$_2$ ambient.
Ni has completely disappeared from the GaN surface, and fully diffused to the top surface. EDX analysis shows that Ni at the top surface is fully oxidized.
On the other hand, Au has fully diffused down to the GaN surface, completely replacing Ni.
This is confirmed by the Z-contrast HAADF-STEM image in the bottom of Figure~\ref{fig:Linescan}: only a weak trace of the initial Ni/Au interface position before annealing is still visible.
Considering the TLM data of Fig.~\ref{fig:TLM}, and Table~\ref{tab:sample-results}, the electrical results obtained after 10 min RTA are slightly improved, but close to that obtained after 2 min RTA.
It is concluded that the absence of Ni, or of NiO$_x$ at the GaN/metal interface, does not degrade the ohmic behaviour of the contact.
Therefore, the presence of Ni/NiO$_x$ at the GaN/metal interface, is not the main origin of the contact improvement, contrary to what was claimed in previous reports\cite{Ho1999,Chen1999}.
Our results rather confirm that an auto-cleaning effect during interdiffusion\cite{Lee1999,Qiao2000,Koide1999}, and/or the creation of Ga vacancies close to the interface\cite{Jang2003,Kim2000,Sarkar2018}, are the most probable origins of the contact improvement.
It should however be noted, that the metal adhesion to the GaN surface has been degraded after 10 min RTA, compared to the 2 min RTA for which no adhesion issue has been detected.
A weak adhesion after important Au interdiffusion has already been reported by Smith et al.\cite{Smith1997}. Practically, a not-too-long annealing time, typically of 2 min to 5 min in our case, seems preferable.

In order to clarify the importance of Ga out-diffusion into Au/Ni and hence the creation of Ga vacancies in the GaN layer, experiments have been repeated with thinner Ni and Au layers in order to improve the resolution of the analysis.
A 3-nm thick Ni layer completed with a 3-nm thick top Au layer has been evaporated onto three additional samples.
The experimental protocol is identical to that presented in the beginning, except for the evaporation rate.
It is reduced to 0.1 nm/s for both Ni and Au in order to get a better thickness control. 

\begin{figure}
\centering{\includegraphics[width=10cm]{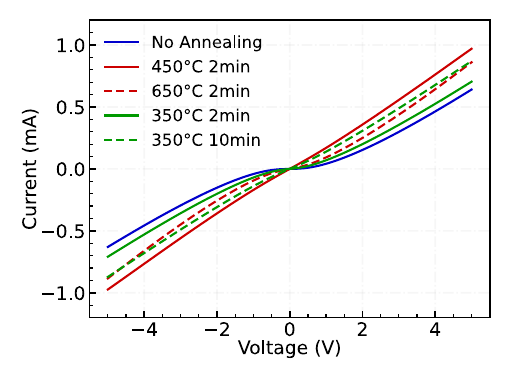}} 
  \caption{
  TLM I-V characteristics measured for a distance between pads L$_i$ = 20 µm without annealing (blue curve), after annealing under O$_2$ atmosphere for 2 min at \SI{450}{\celsius} (red curve), after annealing for under O$_2$ atmosphere for 2 min at \SI{650}{\celsius} (red dashed curve), after annealing under O$_2$ atmosphere at \SI{350}{\celsius} for 2 min (green curve), and for 10 min (green dashed curve). The initial nominal thickness of the Ni, and Au layers evaporated at a rate of 0.1 nm/s is of 3 nm for both layers.}
  \label{fig:TLM_Thin}
\end{figure}

Figure~\ref{fig:TLM_Thin} compares the TLM I-V curves for a distance between TLM pads $L_i$ = 20 µm, for the non-annealed sample, sample after annealing under O$_2$ atmosphere at \SI{450}{\celsius} for 2 min, then at \SI{650}{\celsius} for 2 min, and sample after annealing under O$_2$ atmosphere at \SI{350}{\celsius} for 2 min and 10 min.
The calculated values of $r_c$, $I_{\text{max}}$, and $\Phi_B$ are listed in Table~\ref{tab:sample-results} for each sample.

Comparing the TLM curves of Fig.~\ref{fig:TLM} and Fig.~\ref{fig:TLM_Thin} and the results of Table~\ref{tab:sample-results} for the non-annealed samples with thick, and thin evaporated metal layers, it is first noted that the ohmic behaviour is notably improved with the thin layers.
$\Phi_B$ is decreased by 130 meV, while $I_{\text{max}}$ is increased by a factor of two, and the resulting value of $r_c$ is reduced by a factor of more than 10.
This improvement is attributed to the lower evaporation rate, rather than to the thinner Ni and Au layers, as will be evidenced later from the HR-STEM images and EDX analysis.
Samples with 20-nm, and 200-nm thick Ni, and Au layers deposited at a low-evaporation rate also showed the same improvement of their TLM characteristics before annealing as in Fig.~\ref{fig:TLM_Thin}; however, the detailed analysis and optimization of this effect is beyond the scope of the present study.
Still, RTA under O$_2$ ambient at \SI{450}{\celsius} for 2 min leads to a reduced Schottky barrier height and a higher $I_{\text{max}}$ value as shown in Figure~\ref{fig:TLM_Thin}.
Considering $\Phi_B$ and $I_{\text{max}}$ values, the results after annealing are close to the ones of Fig.~\ref{fig:TLM} obtained for thick metal layers under same annealing conditions.
The Schottky barrier has been apparently completely suppressed in the I-V curve of Fig.~\ref{fig:TLM_Thin} after annealing at \SI{450}{\celsius} for 2 min, leading to the smallest value in Table~\ref{tab:sample-results}.
Although not shown, intermediate annealing tests for 2 min at \SI{500}{\celsius}, and \SI{550}{\celsius} do not modify significantly the TLM characteristics and TLM results.
RTA for 2 min at a high temperature of \SI{650}{\celsius} starts to significantly degrade the contact.
The contact between the metal surface and the electrical probes has been found to be highly resistive and non-reproducible right after the annealing; a second, non-annealed, Ni-Au recharge onto the TLM pads should be evaporated for reproducible and uniform TLM measurements.
The results after recharge evaporation are shown in Fig.~\ref{fig:TLM_Thin}.
Considering the good results already obtained after RTA for 2 min at \SI{450}{\celsius}, it is practically more interesting to verify if a similar improvement can be obtained at a lower thermal budget.
Indeed, as illustrated in Fig.~\ref{fig:TLM_Thin} and in Table~\ref{tab:sample-results}, a low Schottky barrier height and a high $I_{\text{max}}$ value can be obtained at an annealing temperature of \SI{350}{\celsius} for an annealing time of 10 min.
Generally, it can be observed from the results of Table~\ref{tab:sample-results}, that the specific contact resistance is larger for the thin Ni/Au layers, and that for the thick Ni/Au layer.
We attribute this result rather to the more difficult contacting with the electrical probes in the case of thin layers, than to the GaN/metal contact resistance itself.
A metal recharge may improve the retrieved $r_c$ value.

As a practical conclusion from the electrical characterizations, a thin Ni layer completed by an Au layer, annealed at a moderate temperature ($\sim$\SI{350}{\celsius}) under O$_2$ ambient is sufficient to optimize the p-GaN/metal contact, and a low evaporation rate promotes the mechanisms in-play to form the ohmic contact, even without annealing, a result which has not been reported yet for non-alloyed Ni-Au contacts to the best of our knowledge.

\begin{figure}
\centering{\includegraphics[width=\linewidth]{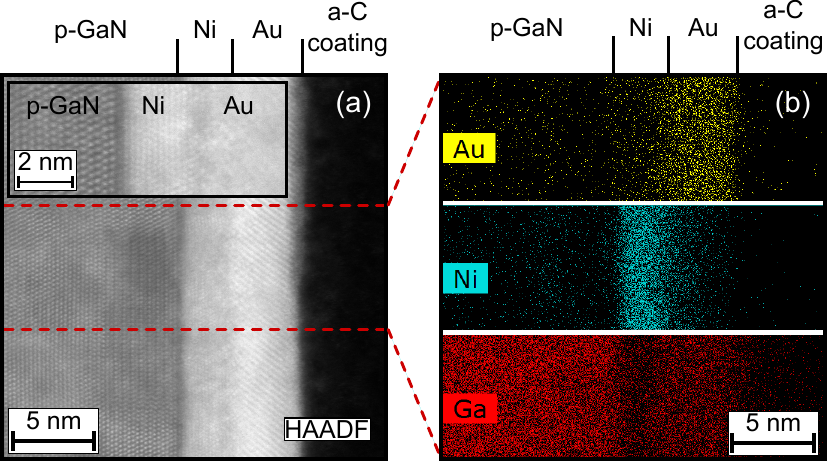}} 
  \caption{
  Cross-sectional HR-STEM image (a) and EDX maps (b) showing the relative distribution of Ga (bottom), Ni (middle), and Au (top) atoms for the non-annealed sample with thin Ni (3nm)/Au (3nm) evaporated layers. The inset in (a) is a higher-magnification HAADF STEM image of the GaN/Ni/Au interfaces. The interval between the horizontal red dashed lines in the HAADF image correspond to the same analysed area in the EDX maps, vertically spaced.}
  \label{fig:Thin_NR}
\end{figure}

Figure~\ref{fig:Thin_NR} shows a cross-sectional HAADF HR-STEM image of the GaN/Ni/Au stack before annealing, and the corresponding EDX maps showing the relative distributions of Ga, Au, and Ni atoms.
The EDX map combining Ga, Ni, and Au relative distributions is given as an insert in Table~\ref{tab:thinNR}.
From the Z-contrast of Fig.~\ref{fig:Thin_NR}-(a), the measured average thickness of the two metal layers is of 2.7 nm ± 0.1 nm, and 2.8 nm ± 0.1 nm, close to the nominal value of 3 nm.
Although not shown, the GaN/Ni interface is found to be uniform at larger-scale and the Ni layer is nano-crystallized as evidenced in the inset of Fig.~\ref{fig:Thin_NR}-(a).
A very thin ($<$ 0.5 nm) layer of lower density (appearing as dark in the HAADF image), presumably an oxide, is only present in localized and discontinuous regions of the GaN/Ni interface.

It could therefore be inferred that this sample is close to the ideal p-GaN/Ni diode for which a moderate Schottky barrier height is theoretically expected\cite{Greco2016,Rickert2002,Wahid2020}, which might explain the better electrical results obtained in comparison with the non-annealed sample with thicker metal layers evaporated at a high rate.
On the other hand, despite the absence of annealing, Fig.~\ref{fig:Thin_NR}-(b) evidenced that a diffusion process has already started to take place.
An interdiffusion between Ni and Au is visible when comparing the top and middle EDX maps.
The most striking result deals with the Ga atoms distribution, that spreads over the initial Ni and Au layers where its relative intensity is much higher than that typically attributed to Ga FIB pollution (visible in the top a-C coating layer).

Following the same approach as in Fig.~\ref{fig:NoAnnealing}-(b), a quantitative analysis of the composition in atomic percentage has been carried out in three rectangular areas selected in the Ni layer, in the Au layer, and in the top of Au layer.
Their position is shown in the inset image of Table~\ref{tab:thinNR}, summarizing the obtained results.
It is observed from the Table that Au-Ni interdiffusion, although moderate, is already present.
The diffusion of Ga into Ni and Au, corresponds to an average Ga composition in atomic percentage in the selected zones \#1 and \#2 above 15\%, and Ga accumulation is evidenced on the top part of the Au layer in zone \#3, where an average Ga composition of 30\% is calculated.
At this extreme surface Gallium is oxidized as deduced from the oxygen percentage. Although not shown, average Ga composition in atomic percentage up to 25\%, and 30\% have been calculated at a different position of the TEM slice, in the Ni layer, and Au layer, respectively.
This level is much higher than the one due to a possible Ga-pollution.
Moreover, in crystallized, Ga-rich, interdiffused Au-Ga zones, the inter-planar spacing (0.333 nm) retrieved from the FFT of TEM diffraction images, is significantly different from inter-distances values expected for pure fcc gold crystal (smaller than 0.240 nm) confirming the presence of Ga in the volume and not only at the slice facets.

\begin{table}
\centering
\caption{Average composition calculated in the zones identified in the inset EDX map corresponding to them TEM image of Figure 7.}
\renewcommand{\arraystretch}{1.8}
\setlength{\tabcolsep}{4pt}
\begin{tabular}{cccccc}
\hline
\makecell{Corresponding\\Figure} & \makecell{Rectangular boxes\\positions located in\\ the EDX map} & \makecell{Element} & \multicolumn{3}{c}{\makecell{Average atomic\\ percentage (\%) in each\\rectangular box \#}} \\
\hline
\multirow{5}{*}{Figure~\ref{fig:Thin_NR}} & \multirow{5}{*}{\raisebox{-0.5\height}{\includegraphics[height=3.5cm]{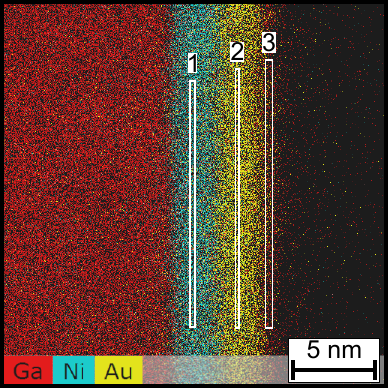}}} & & \#1 & \#2 & \#3 \\
\cline{4-6}
& & Ni & 77.5 & 16.7 & 1.2\\ 
& & Au & 6.1 & 65.1 & 8.8 \\ 
& & Ga & 16.4 & 18.2 & 30.3\\ 
& & O & $\varnothing$ & $\varnothing$ & 59.7\\ 
\hline
\end{tabular}
\par\vspace{0.5em}
The TEM slice thickness is fixed to 200 nm, and the average material density is fixed to 8 g/cm$^3$ for zone \#1 and 15 g/cm$^3$ for zones \#2 and \#3.
\label{tab:thinNR}
\end{table}

The main origin for the onset of the Ni, Au, and Ga diffusion process without any thermal annealing still needs clarification.
Ga diffusion up to the Au layer might have started in-situ during metal evaporation, considering the low evaporation rate and the very thin Ni layer which does not act as an efficient diffusion barrier.
It may also have been ignited in ambient air, after deposition, facilitated by the small thickness ($<$ 3 nm) of the metal layers and assisted by the presence of oxygen.
As an important conclusion of these experimental observations, whatever its origin, the Ga diffusion process must have created Ga vacancies in GaN close to the GaN/metal interface.
It is therefore likely that the Ga vacancies play a crucial role in the improved electrical behaviour of the TLM contacts in Fig.~\ref{fig:TLM_Thin} even before annealing.
Our assumption is consistent with the results of Jang et al.\cite{Jang2003}, showing that the modifications of the GaN surface subsequent to thermal annealing of the Ni/Au contact, and exo-diffusion of Ga into Au, is a crucial mechanism to obtain an ohmic-like p-GaN contact, with any metal\cite{Klump2018}.

\begin{figure}
\centering{\includegraphics[width=\linewidth]{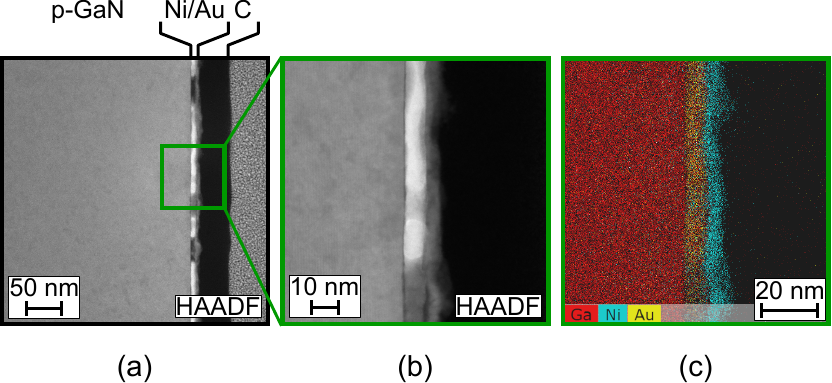}} 
  \caption{Large-scale, cross-sectional HAADF HR-STEM image (a), higher-magnification HAADF STEM image (b), and corresponding EDX map showing the superimposed relative distributions of Ni (blue), Ga (red) and Au (yellow) atoms (c), for a sample with thin Ni (3nm) / Au (3 nm) evaporated layers after thermal annealing at 450 °C for 2 min in O$_2$ atmosphere.}
  \label{fig:Thin_2min}
\end{figure}

Figure~\ref{fig:Thin_2min}-(a) shows a large-scale, HR-STEM cross-sectional image of the GaN/Ni/Au stack at a higher magnification, after RTA under O$_2$ atmosphere at \SI{450}{\celsius} for 2 min.
As reported in Fig.~\ref{fig:TLM_Thin}, and Table~\ref{tab:sample-results}, the electrical contact is improved after such annealing.
The Ni-Au region identified between the GaN surface and the a-C layer coating appears non-uniform at the nano-scale, with some bright and dark areas in the Z-contrast HAADF image.
Dark areas may be attributed to intergranular nano-voids embedded in the TEM slice thickness, formed during the annealing at \SI{450}{\celsius}.
Figure~\ref{fig:Thin_2min}-(b) shows a higher-magnification HAADF HR-STEM image at a frontier between dark/light areas, marked with a green dashed square in Fig.~\ref{fig:Thin_2min}-(a), and the corresponding EDX map with the superimposed relative distributions of Ga, Ni, and Au atoms, used for quantitative analyses of the atomic composition, is shown in Fig.~\ref{fig:Thin_2min}-(c).
The corresponding EDX map with the superimposed relative distributions of Ni and O atoms, showing the three zones selected for the analysis, is reported as an inset in Table~\ref{tab:thin_2min}, summarizing the calculated atomic percentages in the different zones.

It can be concluded from the results of Table~\ref{tab:thin_2min} that, as for the case of 20nm/200nm Ni/Au layers, annealing in O$_2$ atmosphere has promoted the diffusion of Ni on top of the Au layer, where it is fully oxidized with a Ni/O ratio close to 1, while Au has diffused down to the GaN surface replacing Ni, and almost no traces of Au is detected anymore at the top surface.
The Ni-Au inter-diffusion process is almost complete after 2 min at \SI{450}{\celsius}, to be related to the small metal layers thicknesses.
Moreover, TEM observations at highest magnification show that the thin layers are mostly crystallized at the nano-scale, including the NiO$_x$ top layer.
In parallel, Ga atoms out-diffusion into the Au layer has been strongly reinforced during RTA, with more than 60\% of Ga (in atomic percentage) now incorporated, forming an Au-Ga alloy with Ni nano-crystallites, creating more Ga vacancies in the GaN layer.
Finally, it can be noted that the average compositions in atomic percentages retrieved for zone \#1 (selected in a bright region of the Z-contrast HAADF image), and for zone \#2 (selected in a dark region of the Z-contrast HAADF image), are similar, indicating that the change in the intensity is rather due to some nano-domains and nano-voids formed during the annealing at \SI{450}{\celsius} in the thin alloy, and embedded in the thickness of the TEM slice, rather than to a change in composition.

\begin{table}
\centering
\caption{Average composition calculated the rectangular zones identified in the inset EDX map corresponding to TEM image of Figure 8.}
\renewcommand{\arraystretch}{1.8}
\setlength{\tabcolsep}{4pt}
\begin{tabular}{cccccc}
\hline
\makecell{Corresponding\\Figure} & \makecell{Rectangular boxes\\positions located in\\ the EDX map} & \makecell{Element} & \multicolumn{3}{c}{\makecell{Average atomic\\ percentage (\%) in each\\rectangular box \#}} \\
\hline
\multirow{5}{*}{Figure~\ref{fig:Thin_2min}} & \multirow{5}{*}{\raisebox{-0.5\height}{\includegraphics[height=3.5cm]{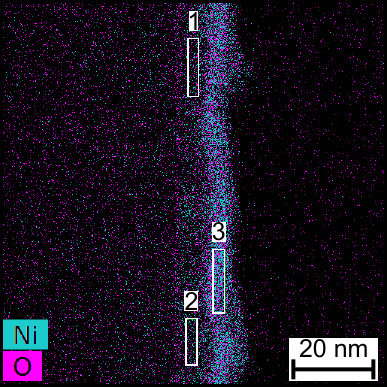}}} & & \#1 & \#2 & \#3 \\
\cline{4-6}
& & Ni & 9.0 & 16.2 & 46.6\\ \cline{3-6}
& & Au & 22.0 & 16.0 & $\varnothing$ \\ \cline{3-6}
& & Ga & 69.0 & 67.8 & 5.1\\ \cline{3-6}
& & O & $\varnothing$ & $\varnothing$ & 48.4\\ 
\hline
\end{tabular}
\par\vspace{0.5em}
The TEM slice thickness is fixed to 100 nm, and the average material density is fixed to 10 g/cm$^3$ for zones \#1 and \#2 and 7 g/cm$^3$ for zone \#3.
\label{tab:thin_2min}
\end{table}

TEM observations and EDX analysis after annealing at \SI{650}{\celsius} for 2 min, showed the same composition, however it was observed that the Au-Ga layer on top of the GaN surface is discontinuous, with large voids clearly apparent on the HAADF images. This may explain the difficult contacting of the TLM pads surface with the electrical probes reported above.

Our STEM-EDX experimental observations, combined with the results of Figs 2 and 6 and of Table~\ref{tab:sample-results}, confirm that the creation of Ga vacancies plays a crucial role in forming an ohmic-like p-GaN/metal contact.
It is also clearly evidenced that neither Ni nor NiO$_x$ is necessarily in contact with p-GaN to obtain such a result.
Rather, a p-GaN/Ga-Au layer seems beneficial to the improvement of the electrical characteristics.
We therefore support the conclusion that creating Ga vacancies represents the key-mechanism to obtain a ohmic contact on p-doped GaN, presumably with any metal.
Our STEM-EDX analysis also explains why removing chemically the NiO$_x$ layer after the RTA step, as proposed by Mengzhe et al.\cite{Mengzhe2009} does not degrade the electrical properties of p-GaN/metal contact.

\section{Conclusion}

We have investigated in detail the formation of an ohmic contact on p-GaN, using a Ni-Au metal layer.
In contrast with previous theoretical considerations reported in literature, our study shows that a Ni/p-GaN interface is not required to obtain the best electrical performance.
We confirm that an O$_2$ atmosphere during post-deposition thermal annealing is determinant to decrease the Schottky barrier height and to improve the contact ohmicity, since oxygen assists the Ni-Au inter-diffusion process during annealing, with Ni being immediately oxidized when reaching the top surface.
More interestingly, our results show that the presence of NiO$_x$ in contact with GaN is not the most important parameter to improve the ohmic behaviour of the contact, in contrast with a recent report.
A GaN/Au-Ga interface leads to better electrical results than a GaN/NiO$_x$ interface.
Indeed, we clearly evidenced that Ga out-diffusion accompanying Au-Ni interdiffusion process to form a Ga-rich, Au-Ga alloy, plays a key-role to create Ga vacancies in the p-GaN.
The best electrical results are obtained in such conditions, and we presume that the creation of Ga vacancies in the top p-GaN surface is the most crucial mechanism in order to obtain an ohmic contact, possibly with any metal consistently with previous reports. 

\section*{Author contributions}
J.D., H.S., M.G., S.B. designed the experiments, fabricated and characterized the chips. G.P. performed the HR-TEM characterization and analysis. D.T. fabricated the FIB slices. T.B. supported the metal evaporation, and N.F. the FIB etching and TEM analysis. P.V., R. J., J.-P. S., T.-M. T., S. S., A. O. designed and grew the epitaxial structure. J.D., H.S., G.P., and S.B analyzed the results, S.B. wrote the article, and prepared the figures with J.D. and H.S. All the authors proof-read the article, and have given approval to the final version of the manuscript.

\section*{Funding}
Part of this work was supported by ANR project CORTIORGAN (ANR-22-CE08-0020-03), and ANR project NEWAVE (ANR-21-CE24-0019-04). C2N and IEMN are members of RENATECH, the national network of large academic micro-nanofabrication facilities.  FIB equipment co-funding: CPER Hauts de France project IMITECH and the Métropole Européenne de Lille.

\section*{Acknowledgments}
The authors thank Francesco Daddi of Grenoble INP Phelma for taking part in the TLM characterizations during his internship in C2N laboratory.

\section*{Abbreviations}
RTA, rapid thermal annealing; EDX, energy dispersive X-ray spectroscopy; HR-TEM, high-resolution transmission electron microscopy; TLM, Transmission Line Method; LED, light-emitting diodes; MQW multi-quantum wells; MOCVD, metal-organic chemical vapor deposition; DIW deionized water; NMP, N-methyl pyrrolidinone; TMAH, tetramethylammonium hydroxide; IPA, isopropyl alcohol. fcc-crystal, face-centered cubic crystal structure.

\bibliographystyle{unsrtnatcustom}
\bibliography{Biblio}


\end{document}